\documentclass[11pt]{article}  
\usepackage[utf8]{inputenc}                

\usepackage[margin=1in]{geometry} 
\usepackage{floatrow}
\usepackage[labelfont={normalsize,sf}, labelformat=simple]{caption, subfig}
\usepackage{mathtools} 
\usepackage{amsthm,amssymb}        
\usepackage{graphicx}
\usepackage{graphbox}
\usepackage{adjustbox}
\usepackage{epstopdf}
\usepackage{array}
\usepackage{xcolor}
\usepackage{breqn}
\usepackage{enumitem}
\usepackage{multirow} 
\usepackage{longtable}
\usepackage{subfiles}
\usepackage{booktabs}
\usepackage{hyperref}
\usepackage[tracking]{microtype}
\usepackage{natbib}
\usepackage{lscape}

\title{A compendium of covariances and correlation coefficients \\ of coalescent tree properties}

\author{Egor Alimpiev\thanks{Department of Biology, Stanford University, Stanford, CA 94305.} \, and Noah A Rosenberg$^*$\\}
\date{\today}

\DeclareMathOperator{\Var}{Var}
\DeclareMathOperator{\Cov}{Cov}
\DeclareMathOperator{\Corr}{Corr}

\newcommand{\expect}[1]{\mathbb{E}\left[#1\right]}
\newcommand{\var}[1]{\Var\left[#1\right]}

\newcommand{\tcov}[1]{\widetilde{\Cov}\left[#1\right]}
\newcommand{\tcorr}[1]{\widetilde{\Corr}\left[#1\right]}

\newcommand{\tcr}[1]{\textcolor{black}{#1}}

\begin{document}

\maketitle

\noindent \small{{\bf Abstract.} Gene genealogies are frequently studied by measuring properties such as their height ($H$), length ($L$), sum of external branches ($E$), sum of internal branches ($I$), and mean of their two basal branches ($B$), and the coalescence times that contribute to the other genealogical features ($T$). These tree properties and their relationships can provide insight into the effects of population-genetic processes on genealogies and genetic sequences. Here, under the coalescent model, we study the 15 correlations among pairs of features of genealogical trees: $H_n$, $L_n$, $E_n$, $I_n$, $B_n$, and $T_k$ for a sample of size $n$, with $2 \leq k \leq n$. We report high correlations among $H_n$, $L_n$, $I_n,$ and $B_n$, with all pairwise correlations of these quantities having values greater than or equal to $\sqrt{6} [6 \zeta(3) + 6 - \pi^2] / ( \pi \sqrt{18 + 9\pi^2 - \pi^4}) \approx 0.84930$ in the limit as $n \rightarrow \infty$, \tcr{where $\zeta$ is the Riemann zeta function}. Although $E_n$ has expectation 2 for all $n$ and $H_n$ has expectation 2 in the $n \rightarrow \infty$ limit, their limiting correlation is 0. The results contribute toward understanding features of the shapes of coalescent trees.}

\clearpage

\section{Introduction}

In coalescent theory, \tcr{features of} gene genealogies are investigated in relation to the evolutionary processes that are included in population-genetic models~\citep{HeinEtAl05, Wakeley09}. For example, comparing a constant-sized and an exponentially growing population, exponential growth increases the total length of the branches of a gene genealogy in relation to its height~\citep{SlatkinAndHudson91, Slatkin96, SanoAndTachida05}. Coalescences are rare in recent generations, when the population is large, and they occur primarily in the period deep in the past when the population was small.

\begin{figure}[htb]
    \centering
    \includegraphics[width=0.50\linewidth]{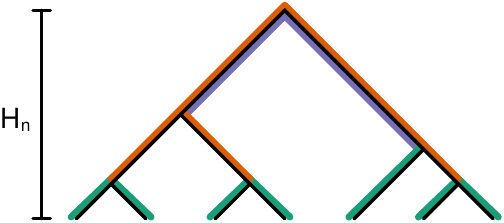}
    \caption{Tree features. The tree height is $H_n$ and the sum of the lengths of all tree branches is $L_n$. External branches, with total length $E_n$, appear in green; internal branches, with total length $I_n$, appear in red; basal branches, with mean length $B_n$, appear in purple.}
    \label{fig:branch_variables}
\end{figure}

Several tree features have been used for measuring effects of population-genetic processes on gene genealogies~\citep{Slatkin96, uyenoyama1997, SchierupAndHein00, Rosenberg06:bookchapter}. For a binary ultrametric tree of $n$ lineages, these features (Figure \ref{fig:branch_variables}) include the tree height from the tips to the root ($H_n$), the total length of all the branches ($L_n$), the total length of external branches connecting tips to the nearest internal node ($E_n$), the total length of internal branches connecting internal nodes to other internal nodes ($I_n$), and the mean length of the two basal branches incident to the root node ($B_n$).

\tcr{These tree features can all be expressed as linear combinations, random linear combinations in some cases, of the same underlying random  variables --- the coalescence times $T_k$ for coalescence of $k$ to $k-1$ lineages, with $2 \leq k \leq n$. Hence, the quantities are correlated. For example, the tree height $H_n$ includes the mean basal branch length $B_n$, and the total length $L_n$ is the sum of the length $E_n$ of the external branches and the length $I_n$ of the internal branches; an increase in $L_n$ necessarily increases $E_n$, $I_n$, or both.}

Analyses of coalescent models have examined some of the correlations between tree features, notably the relationship between $H_n$ and $L_n$~\citep{Fu96, GriffithsAndTavare96, RosenbergAndHirsh03, arbisser2018joint}. Here, we perform a detailed investigation of correlations among $H_n$, $L_n$, $E_n$, $I_n$, and $B_n$. For each pair, under the coalescent, assuming a constant-sized population, we evaluate their covariance and correlation. We explore limiting values as $n \rightarrow \infty$. The approach follows \cite{arbisser2018joint}, who obtained the covariance and correlation of $H_n$ and $L_n$; we perform analogous calculations for all 10 pairs among $\{H_n,L_n,E_n,I_n,B_n\}$, as well as for the five pairs involving one of $\{H_n,L_n,E_n,I_n,B_n\}$ and $T_k$.

\section{Tree properties}
\label{secCoalescentQuantities}

We consider the standard coalescent model of a constant-sized population of size $N$ haploids. Time is measured in units of the population size, with one time unit representing $N$ generations. For sample size $n \geq 2$, we examine tree properties $H_n$, $L_n$, $E_n$, $I_n$, and $B_n$, as well as the coalescence time $T_k$, $2 \leq k \leq n$. In this section, we recall basic features of the various quantities.

For convenience, for a mathematical expression we will use frequently, we write
\begin{equation}
\label{eq:harmonic}
S_{p,n} = \sum_{k=1}^{n}\frac{1}{k^p}.
\end{equation}
\tcr{The limit $S_{p,\infty} = \lim_{n\to\infty} S_{p,n}$ is the Riemann zeta function $\zeta(p)$.} The harmonic sum $S_{1,\infty}$ diverges, and the sum of the reciprocals of squares is $S_{2,\infty}= \pi^2/6 \approx 1.64493$. The sum of the reciprocals of cubes is Ap\'ery's constant, $S_{3,\infty} = \zeta(3) \approx 1.20206$. 

\subsection{$T_k$} 
\label{secTk}
$T_k$ is a random variable representing the time during which $k$ lineages coalesce to $k-1$ lineages. The $T_k$, $2 \leq k \leq n$, are independent and exponentially distributed with probability density function $f_{T_k}(t_k) = \binom{k}{2}e^{-\binom{k}{2}t_k}$~\cite[][p.~60]{Wakeley09}. 
The expectation and variance of $T_k$ are then
\begin{eqnarray}
\label{eq:exp_Tk_def}
\expect{T_k} & = & \frac{2}{k(k-1)}, \\
\label{eq:var_Tk_def}
\var{T_k}     & = & \frac{4}{k^2(k-1)^2}.
\end{eqnarray}
As $n,k \rightarrow \infty$ with $k \leq n$, both $\expect{T_k}$ and $\var{T_k}$ have limit 0.
\subsection{$H_n$}
\label{secHn}
For $n \geq 2$, the height $H_n$ of a tree from root to leaves can be written 
\begin{equation}
\label{eq:Hn_def} 
H_n = \sum_{k=2}^n T_k.
\end{equation}
The expectation and variance of $H_n$ are then found using the expectation and variance of $T_k$ (eqs.~\ref{eq:exp_Tk_def} and \ref{eq:var_Tk_def}), noting that the $T_k$ are independent:
\begin{eqnarray}
\label{eq:exp_Hn_def}
\expect{H_n} & = & \sum_{k=2}^n \expect{T_k} = \frac{2(n-1)}{n}, \\
\label{eq:var_Hn_def}
\var{H_n}    & = & 8\left(\sum_{k=2}^n\frac{1}{k^2}\right) - 4\left(\frac{n-1}{n}\right)^2.
\end{eqnarray}
The variance can be written $\var{H_n}=4(2S_{2,n}n^2-3n^2+2n-1)/n^2$.
The limits are $\lim_{n \rightarrow \infty} \expect{H_n}=2$ and $\lim_{n \rightarrow \infty} \var{H_n} = 4\pi^2/3 - 12 \approx 1.15947$~\cite[][p.~76]{Wakeley09}.

\subsection{$L_n$}
\label{secLn}
For $n \geq 2$, the total tree length, summing the lengths of all branches of a tree, is
\begin{equation}
\label{eq:Ln_def}
L_n = \sum_{k=2}^n kT_k.
\end{equation}
By eqs.~\ref{eq:exp_Tk_def} and \ref{eq:var_Tk_def} and the independence of the $T_k$, we have
\begin{eqnarray}
\label{eq:exp_Ln_def}
\expect{L_n} & = & \sum_{k=2}^n k\expect{T_k} = 2\sum_{k=1}^{n-1} \frac1k, \\
\label{eq:var_Ln_def}
\var{L_n} & = & 4\sum_{k=1}^{n-1} \frac{1}{k^2}.
\end{eqnarray}
In terms of $S_{p,n}$ (eq.~\ref{eq:harmonic}), these expressions are $\expect{L_n}=2S_{1,n-1}$ and $\var{L_n}=4S_{2,n-1}$. The limits are $\lim_{n \rightarrow \infty} \expect{L_n} = \infty$ and $\lim_{n \rightarrow \infty} \var{L_n} = 2\pi^2/3 \approx 6.57974$~\cite[][p.~76]{Wakeley09}.

\subsection{$E_n$}
\label{secEn}
The external branches of a tree are the branches that connect leaves to their nearest internal nodes. Denoting the individual external branch lengths $e_1^{(n)}, e_2^{(n)}, \ldots, e_n^{(n)}$, the sum of external branch lengths is  
\begin{equation*}
E_n = e_1^{(n)} + e_2^{(n)} + \dots + e_n^{(n)}.
\end{equation*}
The $e_k^{(n)}$ are identically distributed, and we write $e_n$ for the length of a randomly chosen external branch of a tree of $n$ lineages. The sum of the external branches has expectation 
\begin{equation}
\label{eq:exp_En}
\expect{E_n}=n\expect{e_n}.
\end{equation}

The random variable $e_n$ can be written recursively as~\citep[][eq.~7]{fu1993statistical} 
\begin{equation}
\label{eq:en_def}
e_n = \begin{cases}
		e_{n-1} + T_n, & \text{\tcr{with} probability } \frac{n-2}{n},\\
		T_n,           & \text{\tcr{with} probability } \frac{2}{n}.\\
\end{cases}
\end{equation}
Expressions for $\expect{e_n}$, $\expect{E_n}$, and $\var{E_n}$ can then be obtained by solving recurrence equations~\citep{fu1993statistical}. We have
\begin{equation}
\label{eq:exp_en_def}
	\expect{e_n} = \frac{2}{n}.
\end{equation}
For the mean and variance of $E_n$, we obtain~\citep[][eqs.~10 and 14]{fu1993statistical}
\begin{eqnarray}
\label{eq:exp_En_def}
	\expect{E_n} & = & 2, \\
\label{eq:var_En_def}
	\var{E_n} &  = & \begin{cases}
		4, & n=2,\\
		\frac{8}{(n-1)(n-2)}[n(\sum_{k=1}^{n-1}\frac{1}{k}) - 2(n-1) ], & n>2.\\
	\end{cases}
\end{eqnarray}
$\expect{E_n}$ is equal to 2 irrespective of the choice of $n$, so that $\lim_{n \rightarrow \infty} \expect{E_n} = 2$. The limit of the variance is $\lim_{n \rightarrow \infty} \var{E_n} = 0$~\citep{fu1993statistical}.

\subsection{$I_n$} 
\label{secIn}
The internal branches connect internal nodes to other internal nodes. Their total length is $I_n$, with 
\begin{equation}
\label{eq:In_def}
	I_n = L_n - E_n. 
\end{equation}
The mean and variance of the sum of internal branches are~\cite[][eqs.~12 and 17]{fu1993statistical}
\begin{eqnarray}
\label{eq:exp_In_def}
\expect{I_n} & = & \expect{L_n} - \expect{E_n} = 2\left(\sum_{k=1}^{n-1}\frac{1}{k}\right) - 2, \\
\var{I_n} & = & 4 \left[\frac{2 [S_{1,n-1}n-2 (n-1)]}{(n-1)(n-2)}-\frac{2 S_{1,n-1}}{n-1}+S_{2,n-1}\right]
	\label{eq:var_In_def}. 
\end{eqnarray}
The limits are $\lim_{n \rightarrow \infty} \expect{I_n} = \infty$ and $\lim_{n \rightarrow \infty} \var{I_n} = 2\pi^2 / 3 \approx 6.57974$, the same as for $L_n$ (Section \ref{secLn}).

\subsection{$B_n$}
\label{secBn}
Finally, we consider the basal branches, the two branches that extend from the root. We define $B_n$ as the mean of the two branch lengths. One of the branches has length $T_2$, and we denote the other length $b_n$. We assume here that $n \geq 4$ for calculations involving $B_n$. The appendix of \cite{uyenoyama1997} gives
\begin{equation}
\label{eq:Bn_def}
	B_n = \frac{T_2 + b_n}{2},
\end{equation}
with 
\begin{equation*}
	b_n = \bigg[ \sum_{j=3}^{n-1}\sum_{k=2}^j\frac{T_k}{j}\prod_{i=3}^{j-1}\left(1-\frac1i\right)\bigg] + \bigg[\sum_{k=2}^n T_k\prod_{i=3}^{n-1}\left(1-\frac1i\right)\bigg]
\end{equation*}
for $n \geq 4$. A convenient form for $b_n$ encodes the fact that with probability $2/[j(j-1)]$, $b_n=H_j$ for $j=3, 4, \ldots, n-1$, and with probability $2/(n-1)$, $b_n=H_n$:
\begin{equation}
\label{eq:bn_convenient}
    b_n = \bigg[ \sum_{j=3}^{n-1} \sum_{k=2}^j \frac{2}{j(j-1)} T_k \bigg] + \bigg( \sum_{k=2}^n \frac{2}{n-1} T_k \bigg).
\end{equation}

Assuming $n \geq 4$, the branch length $b_n$ has expectation~\citep{uyenoyama1997}:
\begin{equation}
\label{eq:exp_bn_def}
	\expect{b_n} = \frac{4}{n} + 4\sum_{k=3}^{n-1}\frac{1}{k^2}.
\end{equation}
The expectation and variance of $B_n$ then equal
\begin{eqnarray}
\label{eq:exp_Bn_def}
\expect{B_n} & = & \frac{2}{n} + 2\sum_{k=2}^{n-1}\frac{1}{k^2}, \\
\label{eq:var_Bn}
\var{B_n} & = & \frac{2(3S_{2,n-1}n^2 - 2S_{2,n-1}^2 n^2 + n^2 - 4S_{2,n-1}n + 3n - 4)}{n^2}.
\end{eqnarray}
The expectation appears in the appendix of \cite{uyenoyama1997}. We calculate the expression for the variance in Section~\ref{calc:Bn}. Taking limits of these equations, we obtain $\lim_{n \rightarrow \infty} \expect{B_n} = \pi^2 / 3 - 2 \approx 1.28987$ and $\lim{n \rightarrow \infty} \var{B_n} = 2+ \pi^2-\pi^4/9 \approx 1.04637$. 

\section{Theoretical results}
\label{secTheoreticalResults}

For pairs of variables among $\{H_n, L_n, E_n, I_n, B_n, T_k\}$, we apply results from Section \ref{secCoalescentQuantities} to compute covariances and correlations. First, for each pair, we compute their covariance. The covariance together with the variances of the two quantities from Section \ref{secCoalescentQuantities} provides their correlation. We obtain the limiting correlation for large trees by taking $n \rightarrow \infty$. Among the 15 pairs, our analyses for 13 are exact; for $(E_n,B_n)$ and $(I_n,B_n)$, we offer approximate covariances and correlations. We also provide the derivation of eq.~\ref{eq:var_Bn} for $\var{B_n}$.

Note that correlations in pairs involving $E_n$ have distinct forms for $n=2$ and $n \geq 3$, owing to the piecewise definition of $\var{E_n}$ in eq.~\ref{eq:var_En_def}. We exclude the case of $n=2$ for pairs involving $I_n$, as $I_2=0$ with $\var{I_2}=0$. We also assume that $B_n$ is defined only for $n \geq 4$.

We present a summary of our mathematical results in Tables \ref{tbl:covariances} and \ref{tbl:correlations}. Table~\ref{tbl:covariances} shows covariances of pairs of variables and their limits as $n \rightarrow \infty$. Table~\ref{tbl:correlations} shows correlations and their $n \rightarrow \infty$ limits.


\subsection{$H_n$ and $T_k$}
\label{calc:HnTk}

We calculate the covariance of $H_n$ and $T_k$ using $\Cov[H_n,T_k] = \expect{H_nT_k}-\expect{H_n}\expect{T_k}$. Recalling that $T_i$ and $T_j$ are independent for $i\neq j$ (Section \ref{secTk}), we have $\expect{T_iT_j} = \expect{T_i}\expect{T_j}$ for $i\neq j$. Hence, inserting eq.~\ref{eq:Hn_def} for $H_n$ and eq.~\ref{eq:exp_Tk_def} for $\expect{T_i}$, we have
\begin{align}
	\Cov[H_n,T_k] & = \expect{T_k\sum_{i=2}^n T_i}-\expect{\sum_{i=2}^n T_i} \expect{T_k}  \nonumber \\
	& = \sum_{i=2}^n \expect{T_iT_k}-\sum_{i=2}^n\expect{T_i}\expect{T_k} \nonumber \\
	& = \var{T_k} + \sum_{i=2,\,i\neq k}^n \expect{T_i}\expect{T_k}-\sum_{i=2,\,i\neq k}^n\expect{T_i}\expect{T_k} = \var{T_k} = \frac{4}{k^2(k-1)^2}, 
\label{eq:CovTkHn} 
\end{align}
where the last step uses $\var{T_k}$ from eq.~\ref{eq:var_Tk_def}. We observe that the covariance is independent of $n$.

For the correlation coefficient $\Corr[H_n,T_k] = \Cov[H_n,T_k]/\sqrt{\var{H_n} \var{T_k}}$, applying eq.~\ref{eq:var_Hn_def} for $\var{H_n}$, eq.~\ref{eq:var_Tk_def} for $\var{T_k}$, and eq.~\ref{eq:CovTkHn} for $\Cov[H_n,T_k]$, we have
\begin{align}
	\Corr[H_n,T_k] 
	& = \frac{n}{\sqrt{2 S_{2,n}n^2  - 3n^2 + 2n - 1 }}\frac{1}{k(k-1)}. \label{eq:CorrTkHn} 
\end{align}
Taking a limit as $n \to \infty$, we obtain
\begin{equation}\label{eq:CorrTkHnlim}
	\lim_{n\to\infty} \Corr[H_n,T_k] = \frac{\sqrt{3}}{\sqrt{\pi^2-9}}\frac{1}{k(k-1)}.
\end{equation}
The limiting correlation decreases to 0 with $k$ from an initial value of $\frac{1}{2}[\sqrt{3/(\pi^2-9)}] \approx 0.92869$ at $k=2$.

\subsection{$L_n$ and $T_k$}
\label{calc:LnTk}

For $L_n$ and $T_k$, applying eqs.~\ref{eq:Ln_def}, \ref{eq:exp_Ln_def}, and \ref{eq:exp_Tk_def}, we have for the covariance 
\begin{align}
	\Cov[L_n,T_k] & = \expect{L_n T_k}-\expect{L_n}\expect{T_k} \nonumber\\
	& = \expect{T_k\sum_{i=2}^n iT_i}-\expect{\sum_{i=2}^n iT_i} \expect{T_k} \nonumber\\
	& = \expect{kT_k^2} - k\expect{T_k}^2 = k\var{T_k} = \frac{4}{k(k-1)^2}, \label{eq:CovTkLn}
\end{align}
where the last step uses eq.~\ref{eq:var_Tk_def}. The covariance of $L_n$ and $T_k$, like $\Cov[H_n,T_k]$ (eq.~\ref{eq:CovTkHn}), is independent of $n$.

Now we calculate the correlation coefficient from eqs.~\ref{eq:CovTkLn}, \ref{eq:var_Ln_def}, and \ref{eq:var_Tk_def}:
\begin{align}
	\Corr[L_n,T_k] 
	& = \frac{1}{\sqrt{S_{2,n-1}}} \frac{1}{k-1}. \label{eq:CorrTkLn}
\end{align}
If we let $n\to\infty$, then this quantity becomes
\begin{equation}\label{eq:CorrTkLnlim}
	\lim_{n\to\infty} \Corr[L_n,T_k] = \frac{\sqrt{6}}{\pi} \frac{1} {k-1}.
\end{equation}
The limiting correlation decreases to 0 with $k$, starting for $k=2$ at $\sqrt{6}/\pi \approx 0.77970$. 

\subsection{$H_n$ and $L_n$}
\label{calc:HnLn}

\cite{arbisser2018joint} reported the covariance and correlation of $H_n$ and $L_n$. By eq.~\ref{eq:Hn_def} and the linearity of the covariance,
\begin{equation}
\Cov[H_n,L_n] = \sum_{k=2}^n \Cov[L_n,T_k]. \nonumber \\
\end{equation}
Applying eq.~\ref{eq:CovTkLn}, we obtain
\begin{equation}
\label{eq:CovHnLn}
\Cov[H_n,L_n] = 4S_{2,n-1}-4+\frac{4}{n}.
\end{equation}
The limit of the covariance is 
\begin{equation}
\label{eq:CovHnLnlim}
\lim_{n \rightarrow \infty} \Cov[H_n,L_n] = \frac{2\pi^2}{3} - 4 \approx 2.57974.
\end{equation}

Dividing the covariance in eq.~\ref{eq:CovHnLn} by the square root of the product of eqs.~\ref{eq:var_Hn_def} and \ref{eq:var_Ln_def}, we obtain
\begin{equation}
\label{eq:corr_Hn_Ln}
\Corr[H_n,L_n] = \frac{S_{2,n-1}n-n+1}{\sqrt{S_{2,n-1} (2 S_{2,n} n^2- 3n^2 + 2n -1 )}}.
\end{equation}
The limit is 
\begin{equation}
    \lim_{n \rightarrow \infty} \Corr[H_n,L_n] = \frac{\pi^2-6}{\pi \sqrt{2(\pi^2-9)}} \approx 0.93399.
\label{eq:corr_Hn_Ln_limit}
\end{equation}
\subsection{$H_n$ and $E_n$}
\label{calc:HnEn}

For the covariance $\Cov[H_n,E_n] = \expect{H_nE_n} - \expect{H_n}\expect{E_n}$, we first note that by eqs.~\ref{eq:exp_Hn_def} and \ref{eq:exp_En_def}, the second term is simply $4\left(1-\frac1n\right)$. Expanding $\expect{H_n E_n}$ by using the definition of $H_n$ (eq.~\ref{eq:Hn_def}) gives us
\begin{equation*}
	\expect{H_n E_n} = \sum_{i=2}^n \expect{E_nT_i} = n\sum_{i=2}^n \expect{e_nT_i},
\end{equation*}
as all external branch lengths are identically distributed (eq.~\ref{eq:exp_En}). 

For integers $k$, $i$ with $2 \leq k, i, \leq n$, the external branch length $e_k$, representing the length of a randomly chosen external branch for a tree with $k$ leaves, and the coalescence time $T_i$, satisfy (eq.~\ref{eq:en_def})
\begin{equation}
\label{eq:ekTi}
	e_kT_i = \begin{cases}
		e_{k-1}T_i + T_kT_i, & \text{\tcr{with} probability } \frac{k-2}{k},\\
		T_kT_i, & \text{\tcr{with} probability } \frac{2}{k},\\
	\end{cases}
\end{equation}
where for convenience, we write $e_1=0$.

Note that $e_k$ and $T_i$ are independent for $i > k$, as the recurrence for $e_k$ is constructed only using coalescence times $T_2, T_3, \ldots, T_k$ (eq.~\ref{eq:en_def}); each of these times is indepdendent of $T_i$ for $i > k$ (Section \ref{secTk}). We solve to find $\expect{e_nT_i}$ by computing $\expect{e_kT_i}$, incrementing $k$ from 2 to $n$. The calculations are similar to those of the Appendix of  \cite{fu1993statistical}.

$\expect{e_2T_2}$ is trivial, with $e_2=T_2$, and $\expect{e_2T_2}=\expect{T_2^2}=2$ by eqs.~\ref{eq:exp_Tk_def} and \ref{eq:var_Tk_def}. By eqs.~\ref{eq:exp_en_def} and \ref{eq:exp_Tk_def} and the independence of $e_k$ and $T_i$ for $i > k$, for $i \geq 3$,
\begin{equation*}
\expect{e_{i-1}T_i} = \expect{e_{i-1}}\expect{T_i} = \frac{4}{i(i-1)^2}.
\end{equation*}
Noting $\expect{T_i^2} = \var{T_i} + \expect{T_i}^2 = 2\expect{T_i}^2$ by eqs.~\ref{eq:exp_Tk_def} and \ref{eq:var_Tk_def}, we use eq.~\ref{eq:ekTi} to write an expression for $\expect{e_iT_i}$:
\begin{equation}
	\expect{e_iT_i}  = \frac{i-2}{i}\expect{e_{i-1}T_i} + \expect{T_i^2} 
					= \frac{4}{i(i-1)^2}. \label{eq:exp_eiTi}
\end{equation}

Next, incrementing eq.~\ref{eq:exp_eiTi}, we have
\begin{equation}
	\expect{e_{i+1}T_i} = \frac{i-1}{i+1}\expect{e_iT_i} + \expect{T_{i+1}}\expect{T_i}
						= \frac{4}{i^2(i-1)},
\label{eq:exp_enTi_initial}
\end{equation}
by eqs.~\ref{eq:exp_Tk_def} and \ref{eq:exp_en_def}.

The final step is to solve the recurrence equation
\begin{equation*}
	\expect{e_nT_i} = \frac{n-2}{n}\expect{e_{n-1}T_i}+\expect{T_n}\expect{T_i},
\end{equation*}
with initial condition eq.~\ref{eq:exp_enTi_initial}. Recalling the case of $i=n=2$, with $2 \leq i \leq n$, we obtain solution
\begin{equation}
\label{eq:exp_enTi}
	\expect{e_nT_i} = \frac{4}{i(i-1)(n-1)}.
\end{equation}

Applying eq.~\ref{eq:exp_En}, the expression for $\Cov[H_n,E_n]$ becomes
\begin{equation}
	\Cov[H_n,E_n] = n\sum_{i=2}^n\frac{4}{i(i-1)(n-1)}-4\left(1-\frac1n\right) 
	= \frac{4}{n}.
	\label{eq:CovHnEn}
\end{equation}
The limit of the covariance as $n \rightarrow \infty$ is  
\begin{equation}\label{eq:CovHnEnlim}
\lim_{n \to \infty} \Cov[H_n,E_n] = 0.
\end{equation}

Dividing eq.~\ref{eq:CovHnEn} by the square root of the product of variances from eqs.~\ref{eq:var_Hn_def} and \ref{eq:var_En_def}, the correlation is
\begin{align}
	\Corr[H_n,E_n] = 
\begin{cases}
1, & n=2, \\
\frac{\sqrt{(n-1) (n-2)}}{\sqrt{2 \left(2 S_{2,n}n^2 - 3n^2 + 2n - 1 \right) \left( S_{1,n-1}n-2 n+2\right)}}, & n > 2.
\end{cases}
\label{eq:CorrHnEn}
\end{align}
The limit of the correlation is  
\begin{equation}\label{eq:CorrHnEnlim}
	\lim_{n\to\infty} \Corr[H_n,E_n] = 0.
\end{equation}

\subsection{$E_n$ and $T_k$}
\label{calc:EnTk}

In the process of computing $\Cov[H_n,E_n]$, we have obtained an expression for $\expect{e_nT_i}$ (eq.~\ref{eq:exp_enTi}), from which we can obtain  $\Cov[E_n,T_k] = n\expect{e_nT_k} - \expect{E_n}\expect{T_k}$. Applying eqs.~\ref{eq:exp_En_def} and \ref{eq:exp_Tk_def}, we have
\begin{align}
    \Cov[E_n,T_k] 
                  & = \frac{4}{k(k-1) (n-1)}. \label{eq:CovEnTk}
\end{align}
Irrespective of the value of $k$, we have 
\begin{equation}\label{eq:CovEnTklim}
	\lim_{n\to \infty} \Cov[E_n,T_k] =0.
\end{equation}

Applying eqs.~\ref{eq:var_Tk_def} and \ref{eq:var_En_def}, the correlation coefficient is 
\begin{align}\label{eq:CorrEnTk}
	\Corr[E_n,T_k] & = \begin{cases}
	1, & n=2, \\
	\frac{\sqrt{n-2}}{\sqrt{2(n-1)(S_{1,n-1}n-2 n+2)}}, & n \geq 3.
	\end{cases}
\end{align}
The correlation coefficient is independent of $k$, and it has limit
\begin{equation}\label{eq:CorrEnTklim}
	\lim_{n\to\infty} \Corr[E_n,T_k] = 0.
\end{equation}

\subsection{$L_n$ and $E_n$}
\label{calc:LnEn}

\cite{fu1993statistical} provided the expression for $\expect{L_nE_n}$ (see also p.\ 167 of \cite{durrett}, with all values scaled by $2N_e$). The main result is the following expression, obtained by solving recurrence equations:
\begin{equation*}
	\expect{L_nE_n} = \frac{4S_{1,n-1} n}{n-1}.
\end{equation*}
We can use this result to calculate the covariance of $L_n$ and $E_n$ by $\Cov[L_n,E_n]=\expect{L_nE_n} - \expect{L_n} \expect{E_n}$ with eqs.~\ref{eq:exp_Ln_def} and \ref{eq:exp_En_def}. The covariance can also be quickly obtained from eqs.~\ref{eq:Ln_def} and \ref{eq:CovEnTk},
\begin{equation}
\Cov[L_n,E_n]  = \sum_{k=2}^n k \Cov[E_n,T_k]
%
			   = \frac{4S_{1,n-1}}{n-1}. \label{eq:CovLnEn}
\end{equation}
The limit is
\begin{equation}
	\label{eq:CovLnEnlim}
	\lim_{n\to\infty} \Cov[L_n,E_n] = 0.
\end{equation}

Applying eqs.~\ref{eq:CovLnEn}, \ref{eq:var_Ln_def}, and \ref{eq:var_En_def}, the correlation coefficient of $L_n$ and $E_n$ is  \begin{align}
	\Corr[L_n,E_n] & = 
				   \begin{cases}
				   1, &  n=2, \\
				   \frac{S_{1,n-1} \sqrt{n-2} }{  \sqrt{2S_{2,n-1}(n-1) (S_{1,n-1}n-2 n+2)}}, &  n \geq 3,
				   \end{cases}  \label{eq:CorrLnEn}
\end{align}
with the limit
\begin{equation}
\label{eq:CorrLnEnlim}
	\lim_{n\to\infty}\Corr[L_n,E_n]=0. 
\end{equation}
			  
\subsection{$H_n$ and $I_n$}
\label{calc:HnIn}

For the pair $H_n$ and $I_n$, we exploit results obtained for other pairs to quickly obtain the covariance. Remembering that $I_n = L_n - E_n$ (eq.~\ref{eq:In_def}), we use eqs.~\ref{eq:CovHnLn} and \ref{eq:CovHnEn} to obtain for $n \geq 3$
\begin{align}
	\Cov[H_n,I_n] & = \Cov[H_n,L_n] - \Cov[H_n,E_n] \nonumber \\
	 & = 4S_{2,n-1} - 4. \label{eq:CovHnIn} 
\end{align}
For this covariance, we have 
\begin{equation}\label{eq:CovHnInlim}
	\lim_{n\to\infty} \Cov[H_n,I_n] = \frac{2\pi^2}{3}-4 \approx 2.57974. 
\end{equation}

From the covariance in eq.~\ref{eq:CovHnIn} and variances in eqs.~\ref{eq:var_Hn_def} and \ref{eq:var_In_def}, we compute the correlation coefficient:
\begin{equation}
	\label{eq:CorrHnIn}
\Corr[H_n,I_n] = 
\frac{\left(S_{2,n-1}-1\right)n \sqrt{(n-1)(n-2)} } {\sqrt{\left(2 S_{2,n}n^2-3 n^2+2 n-1\right) \left[4 S_{1,n-1}+S_{2,n-1}(n-1)(n-2) -4(n-1)\right]}}.
\end{equation}
The limit is the same as that of $\Corr[H_n,L_n]$, or (eq.~\ref{eq:corr_Hn_Ln_limit}) 
\begin{equation}
	\label{eq:CorrHnInlim}
	\lim_{n\to\infty} \Corr[H_n,I_n] = \frac{\pi ^2-6}{\pi  \sqrt{2 \left(\pi ^2-9\right)}} \approx 0.93399.
\end{equation}

\subsection{$I_n$ and $T_k$}
\label{calc:InTk}

By eqs.~\ref{eq:In_def}, \ref{eq:CovTkLn}, and \ref{eq:CovEnTk}, assuming $n \geq 3$, we have
\begin{align}
    \Cov[I_n,T_k] & = \Cov[L_n,T_k] - \Cov[E_n,T_k] \nonumber \\ 
    & = \frac{4 (n-k)}{k(k-1)^2 (n-1)}. \label{eq:CovInTk}
\end{align}
The limit of this expression is a rapidly decreasing function of $k$,
\begin{equation}
	\label{eq:CovInTklim}
	\lim_{n\to\infty} \Cov[I_n,T_k] = \frac{4}{k(k-1)^2}.
\end{equation}

Using the variances in eqs.~\ref{eq:var_In_def} and \ref{eq:var_Tk_def}, the correlation coefficient is
\begin{equation}
	\label{eq:CorrInTk}
	\Corr[I_n,T_k] =\frac{(n-k)\sqrt{n-2}}{(k-1) \sqrt{(n-1) \left[4 S_{1,n-1} +  S_{2,n-1} (n-1)(n-2)-4(n-1) \right]} },
\end{equation}
with limit
\begin{equation}
	\label{eq:CorrInTklim}
	\lim_{n\to\infty} \Corr[I_n,T_k] = \frac{\sqrt{6}}{\pi} \frac{1}{k-1}.
\end{equation}
The limit \tcr{of $\Corr[I_n,T_k]$ as $n\to\infty$ is equal to} that of $\Corr[L_n,T_k]$ (eq.~\ref{eq:CorrTkLnlim}),

\subsection{$L_n$ and $I_n$}
\label{calc:LnIn}

By eq.~\ref{eq:In_def}, we can apply eqs.~\ref{eq:var_Ln_def} and \ref{eq:CovLnEn} to obtain for $n \geq 3$
\begin{align}
	\Cov[L_n,I_n] & = \var{L_n} - \Cov[L_n,E_n] \nonumber \\
	& = 4 S_{2,n-1}-\frac{4 S_{1,n-1}}{n-1}. \label{eq:CovLnIn}
\end{align}
The limit as $n \to \infty$ is
\begin{equation}\label{eq:CovLnInlim}
	\lim_{n\to\infty} \Cov[L_n,I_n] = \frac{2 \pi ^2}{3} \approx 6.57974.
\end{equation}

For the correlation coefficient, applying eqs.~\ref{eq:CovLnIn}, \ref{eq:var_Ln_def}, and \ref{eq:var_In_def}, we get
\begin{equation}
	\label{eq:CorrLnIn}
	\Corr[L_n,I_n] = \frac{[S_{2,n-1}(n-1) -S_{1,n-1}]\sqrt{n-2}}{(n-1) \sqrt{S_{2,n-1} \left[4 S_{1,n-1}+S_{2,n-1}(n-1)(n-2) -4(n-1)\right]}},
\end{equation}
with 
\begin{equation}\label{eq:CorrLnInlim}
	\lim_{n\to\infty} \Corr[L_n,I_n] = 1.
\end{equation}

\subsection{$E_n$ and $I_n$}
\label{calc:EnIn}

For this pair, with $n \geq 3$, the covariance was reported by  \cite{fu1993statistical}:
\begin{equation}\label{eq:CovEnIn}
	\Cov[E_n,I_n] = \frac{4S_{1,n-1}}{n-1}- \frac{8S_{1,n-1}n}{(n-1)(n-2)} + \frac{16}{n-2}.
\end{equation}
We can also obtain this result quickly from eqs.~\ref{eq:In_def}, \ref{eq:CovLnIn}, and \ref{eq:var_In_def}, as $\Cov[E_n,I_n]=\Cov[L_n-I_n,I_n] = \Cov[L_n,I_n]- \Var[I_n]$.
In the limit, we have 
\begin{equation}\label{eq:CovEnInlim}
	\lim_{n\to\infty} \Cov[E_n,I_n] = 0.
\end{equation}

For the correlation coefficient, we divide eq.~\ref{eq:CovEnIn} by the product of the square roots of eqs.~\ref{eq:var_En_def} and \ref{eq:var_In_def}:
\begin{equation}\label{eq:CorrEnIn}
	\Corr[E_n,I_n] = \frac{4 (n-1)-S_{1,n-1}(n+2)}{\sqrt{2\left( S_{1,n-1}n-2 n+2\right) \left[4 S_{1,n-1}+S_{2,n-1}(n-1)(n-2) -4(n-1)\right]}},
\end{equation}
with the limit
\begin{equation}\label{eq:CorrEnInlim}
	\lim_{n\to\infty} \Corr[E_n,I_n] = 0.
\end{equation}
This result \tcr{is equal to} the limit for $\Corr[L_n, E_n]$ (eq.~\ref{eq:CorrLnEnlim}).

\subsection{$\var{B_n}$}
\label{calc:Bn}

To obtain correlation coefficients involving $B_n$, assuming $n \geq 4$, we first verify the expression for $\var{B_n}$ in eq.~\ref{eq:var_Bn}. By definition of $B_n$ in eq.~\ref{eq:Bn_def}, we have
\begin{equation}
	\var{B_n} = \expect{\frac{1}{4}(T_2 + b_n)^2} - \expect{\frac{1}{2}(T_2 + b_n)}^2 = \frac{1}{4} \var{b_n} + \frac{1}{2} \Cov[b_n, T_2] + \frac{1}{4},
\label{eq:VarBnFirstStep}
\end{equation}
where we have used $\var{T_2} = 1$ (eq.~\ref{eq:var_Tk_def}). 

To calculate $\var{b_n}$, we first recall that for $j=3, 4, \ldots, n-1$, with probability $2/[j(j-1)]$, we have $b_n=H_j$; with probability $2/(n-1)$ we have $b_n=H_n$. Hence, applying eq.~\ref{eq:bn_convenient} and $\expect{H_k^2}=\var{H_k}+\expect{H_k}^2$ with eqs.~\ref{eq:exp_Hn_def} and \ref{eq:var_Hn_def}, we have
\begin{align*}
	\expect{b_n^2} & = \bigg[ \sum_{j = 3}^{n-1} \frac{2}{j(j-1)} \expect{H_j}^2 \bigg] + \frac{2}{n-2} \expect{H_n}^2 \\
& = \frac{-16S_{2,n-1}n^2+30n^2 - 16n - 16}{n^2}.
\end{align*}
Using the expression for $\expect{b_n}$ from eq.~\ref{eq:exp_bn_def}, we obtain 
\begin{equation}
\label{eq:Varbn}
\var{b_n} = \frac{24S_{2,n-1}n^2 - 16S_{2,n-1}^2 n^2 + 5n^2 -32S_{2,n-1}n+24n - 32}{n^2}.
\end{equation}

Next, we compute $\Cov[b_n,T_k]$ and insert $k=2$. By eq.~\ref{eq:bn_convenient}, applying the independence of the $T_i$ (Section \ref{secTk}) and inserting eq.~\ref{eq:var_Tk_def}, we have
\begin{align}
    \Cov[b_n,T_k] & = \bigg[ \sum_{j=3}^{n-1} \sum_{i=2}^j \frac{2}{j(j-1)} \Cov[T_i,T_k] \bigg] + \bigg( \sum_{i=2}^n \frac{2}{n-1} \Cov[T_i,T_k]\bigg)  \nonumber \\
    & = \bigg[ \sum_{i=2}^{n-1} \sum_{j=i \text{ if } i \geq 3 \atop j=3 \text{ if } i=2}^{n-1} \frac{2}{j(j-1)} \Cov[T_i,T_k] \bigg] + \bigg( \frac{2}{n-1} \var{T_k} \bigg) \nonumber \\
    & = 
    \begin{cases}
    \left[ \sum_{j=3}^{n-1} \frac{2}{j(j-1)} \var{T_k} \right] + \left( \frac{2}{n-1}\var{T_k} \right), & k=2, \\
    \left[ \sum_{j=k}^{n-1} \frac{2}{j(j-1)} \var{T_k} \right] + \left( \frac{2}{n-1}\var{T_k} \right), &  k=3,4,\ldots,n-1, \\ 
    \frac{2}{n-1}\var{T_k}, & k=n,
    \end{cases} \nonumber \\
    & = 
    \begin{cases}
    1, & k=2, \\
    \frac{8}{k^2(k-1)^3}, & k=3, 4, \ldots, n.
    \end{cases} \label{eq:covbnTk}
\end{align}
Inserting $\var{b_n}$ from eq.~\ref{eq:Varbn} and $\Cov[b_n,T_2]$ from eq.~\ref{eq:covbnTk} into eq.~\ref{eq:VarBnFirstStep}, we confirm eq.~\ref{eq:var_Bn}.
\subsection{$B_n$ and $T_k$}
\label{calc:BnTk}

We extract $\Cov[B_n,T_k]$ from Section~\ref{calc:Bn}, as $\Cov[B_n,T_k] = {\Cov[b_n,T_k]}/{2} + {\Cov[T_2,T_k]}/{2}$ by the definition in eq.~\ref{eq:Bn_def}, and $\Cov[T_2,T_k]=\delta_{k,2}$, where $\delta$ is the Kronecker delta (Section \ref{secTk}). By eq.~\ref{eq:covbnTk}, recalling $n \geq 4$,
\begin{equation}
\Cov[B_n,T_k] = \frac{4}{k^2(k-1)^3}.
    \label{eq:CovBnTk}
\end{equation}
The covariance is independent of $n$.

For the correlation coefficient, using eqs.~\ref{eq:CovBnTk}, \ref{eq:var_Bn}, and \ref{eq:var_Tk_def}, we have
\begin{equation}
	\label{eq:CorrBnTk}
	\Corr[B_n,T_k]  = 
	\frac{\sqrt{2}n}{\sqrt{3S_{2,n-1}n^2 - 2S_{2,n-1}^2 n^2 + n^2 -4S_{2,n-1}n +3n-4}} \frac{1}{k(k-1)^2},
\end{equation}
with the limit
\begin{equation}
	\label{eq:CorrBnTklim}
	\lim_{n\to\infty} \Corr[B_n,T_k] = \frac{6}{\sqrt{18+9 \pi ^2-\pi ^4}} \frac{1}{k(k-1)^2}.
\end{equation}
The limit begins at $3/\sqrt{18+9 \pi ^2-\pi ^4} \approx 0.97759$ for $k=2$ and quickly decreases to 0 as $k$ increases.

\subsection{$H_n$ and $B_n$}
\label{calc:HnBn}

To obtain $\Cov[H_n,B_n]$ with $n \geq 4$, we begin from eq.~\ref{eq:Bn_def}:
\begin{equation*}
	\Cov[H_n,B_n] = \frac{1}{2}\Cov[H_n,b_n] + \frac{1}{2}\Cov[H_n,T_2].
\end{equation*}
The second term was computed in eq.~\ref{eq:CovTkHn}, $\Cov[H_n,T_2]=1$.

For the first term, $\Cov[H_n,b_n]$, we decompose $H_n$ (eq.~\ref{eq:Hn_def}) and apply eq.~\ref{eq:covbnTk} to obtain
\begin{align}
    \Cov[H_n,b_n] & = \sum_{k=2}^n \Cov[b_n,T_k] \nonumber \\
    & = 1 + \sum_{k=3}^n \frac{8}{k^2 (k-1)^3}. \nonumber 
\end{align}
We use a partial fraction decomposition to sum the series, obtaining 
\begin{align}
    \Cov[H_n,B_n] & = 1 + \sum_{k=3}^n \frac{4}{k^2 (k-1)^3} \nonumber \\ & = \frac{4[S_{3,n-1}n^2 - 3S_{2,n-1}n^2 + (n-1)(4n+1)]}{n^2}.
\label{eq:CovHnBn}
\end{align}

The asymptotic limit of $\Cov[H_n,B_n]$ is
\begin{equation}\label{eq:CovHnBnlim}
	\lim_{n\to\infty} \Cov[H_n,B_n] = 4 \zeta (3) + 16 - 2 \pi ^2 \approx 1.06902.
\end{equation}

The correlation coefficient is then equal to:
\begin{equation}
	\label{eq:CorrHnBn}
\Corr[H_n,B_n] = \frac{\sqrt{2}[S_{3,n-1}n^2 - 3S_{2,n-1}n^2 + (n-1)(4n+1)]}{ \sqrt{(2S_{2,n}n^2-3n^2+2n-1) (3S_{2,n-1}n^2-2S_{2,n-1}^2n^2+ n^2-4S_{2,n-1}n+3n-4})}.
\end{equation}

The limit of the correlation coefficient is:
\begin{equation}
	\label{eq:CorrHnBnlim}
	\lim_{n\to\infty} \Corr[H_n,B_n] = \frac{3\sqrt{3} [2\zeta(3) +8 -\pi^2]}{\sqrt{(\pi^2 - 9)(18+9\pi^2-\pi^4)}} \approx 0.97054.
\end{equation}

\subsection{$L_n$ and $B_n$}
\label{calc:LnBn}

In a manner similar to that used in Section~\ref{calc:HnBn}, with $n \geq 4$, we expand $\Cov[L_n,B_n]$ using eq.~\ref{eq:Bn_def}:
\begin{equation*}
	\Cov[L_n,B_n] = \frac{1}{2}\Cov[L_n,b_n] + \frac{1}{2}\Cov[L_n,T_2].
\end{equation*}
The second term is $\Cov[L_n,T_2]=2$ by eq.~\ref{eq:CovTkLn}. The first term is decomposable by eq.~\ref{eq:Ln_def}; applying eq.~\ref{eq:covbnTk}, 
\begin{align}
    \Cov[L_n,b_n] & = \sum_{k=2}^n k \Cov[b_n,T_k] \nonumber \\
    & = 2 + \sum_{k=3}^n \frac{8}{k(k-1)^3}. \nonumber
\end{align}
Summing the series, we have
\begin{align}
    \Cov[L_n,B_n] & = 2 + \sum_{k=3}^n \frac{4}{k(k-1)^3} \nonumber \\
& = \frac{4[S_{3,n-1}n - S_{2,n-1}n + n-1]}{n}.
\label{eq:CovLnBn}
\end{align}

The limiting covariance is 
\begin{equation}\label{eq:CovLnBnlim}
	\lim_{n\to\infty}\Cov[L_n,B_n] = 4\zeta(3) + 4 - \frac{2\pi^2}{3} 
\approx 2.22849.
\end{equation}

Using eqs.~\ref{eq:CovLnBn}, \ref{eq:var_Ln_def}, and \ref{eq:var_Bn}, we now obtain an expression for the correlation coefficient:
\begin{equation}\label{eq:CorrLnBn}
\Corr[L_n,B_n] = \frac{\sqrt{2}(S_{3,n-1}n - S_{2,n-1}n + n-1)} { \sqrt{S_{2,n-1}({3S_{2,n-1}n^2-2S_{2,n-1}^2n^2+n^2-4S_{2,n-1}n+3n-4})}}.
\end{equation}
The limit is 
\begin{equation}\label{eq:CorrLnBnlim}
	\lim_{n\to\infty} \Corr[L_n,B_n] = \frac{\sqrt{6} [6\zeta(3) + 6 - \pi^2]}{\pi \sqrt{18+9\pi^2-\pi^4} } \approx 0.84930.
\end{equation}


\subsection{$E_n$ and $B_n$}
\label{calc:EnBn}

For $\Cov[E_n,B_n]$, we obtain an approximate rather than exact answer. Decomposing $B_n$ by eq.~\ref{eq:Bn_def}, we have
\begin{equation}
\label{eq:covEnBn1}
\Cov[E_n,B_n] = \frac{1}{2}\Cov[E_n,T_2] + \frac{1}{2} \Cov[E_n,b_n].
\end{equation}

Recall that $b_n$ can be defined conditionally, \tcr{in terms of a random variable $J$ that characterizes the coalescence times that it contains (Section~\ref{secBn}).} \tcr{More precisely, we say that f}or a random variable $J$, $b_n=H_j$ with probability $p_j$, where $p_j=\mathbb{P}[J=j]=2/[j(j-1)]$ for $J=3,4,\ldots,n-1$ and $p_j=2/(j-1)$ for $J=n$. We can then decompose the covariance $\Cov[E_n,b_n]$ by the conditional covariance formula, conditioning on $J$:
\begin{equation}
\label{eq:covEnbn2}
\Cov[E_n,b_n] = \mathbb{E} \bigg[ \Cov[E_n,b_n | J] \bigg] + \Cov \bigg[\expect{E_n|J} , \expect{B_n | J} \bigg].
\end{equation}

We next perform an approximation by ignoring the second term in the covariance decomposition. Noting that 
$\Cov[E_n,T_2]=\frac{2}{n-1}$ by eq.~\ref{eq:CovEnTk}, we use eq.~\ref{eq:covEnBn1} together with eq.~\ref{eq:covEnbn2} to write approximations
\begin{align}
\label{eq:approxCovEnbn1}
\tcov{E_n,b_n} & = \mathbb{E} \bigg[ \Cov[E_n,b_n | J] \bigg] \\
\label{eq:approxCovEnBn}
\tcov{E_n,B_n} & = \frac{1}{n-1} + \frac{1}{2} \tcov{E_n,b_n}.
\end{align}

Weighting each $\Cov[H_j,E_n]$ by the associated probability $p_j$, and decomposing $H_j$ by eq.~\ref{eq:Hn_def}, eq.~\ref{eq:approxCovEnbn1} gives 
\begin{align}
\mathbb{E} \bigg[ \Cov[E_n,b_n | J] \bigg] & =  \sum_{j=3}^n p_j \Cov[E_n,b_n | J=j] \nonumber \\
& =  \bigg[ \sum_{j=3}^{n-1} \frac{2}{j(j-1)} \Cov[H_j,E_n] \bigg] + \frac{2}{n-1} \Cov[H_n,E_n] \nonumber \\
& = \bigg[ \sum_{j=3}^{n-1} \frac{2}{j(j-1)} \sum_{k=2}^j \Cov[E_n,T_k] \bigg] + \frac{2}{n-1} \sum_{k=2}^j \Cov[E_n,T_k].
\nonumber
\end{align}
We can then insert the result of eq.~\ref{eq:CovEnTk} and simplify to obtain
\begin{equation}
\label{eq:tcovEnbn}
    \tcov{E_n,b_n} = \frac{2 (4 S_{2, n-1} n -5 n+4)}{n(n-1)}.
\end{equation}

Finally, inserting eq.~\ref{eq:tcovEnbn} into eq.~\ref{eq:approxCovEnBn}, we have 
\begin{equation}
\label{eqCovEnBn}
    \tcov{E_n,B_n} = \frac{4 (S_{2, n-1}n-n+1)}{n(n-1) }. 
\end{equation}
The limit is 
\begin{equation}
    \lim_{n \rightarrow \infty} \tcov{E_n,B_n} = 0. 
\label{eqCovEnBnLim}
\end{equation}

For the approximate correlation coefficient $\tcorr{E_n,B_n} = \tcov{E_n,B_n} / \sqrt{\var{E_n} \var{B_n}}$, we use eqs.~\ref{eqCovEnBn}, \ref{eq:var_En_def}, and \ref{eq:var_Bn} to obtain 
\begin{equation}
    \tcorr{E_n,B_n} = \frac{(S_{2, n-1}n-n+1)\sqrt{n-2}}
    {\sqrt{ (n-1) (S_{1, n-1}n -2 n+2) (3S_{2, n-1}n^2 -2S_{2,n-1}^2 n^2 +n^2 - 4S_{2, n-1} n +3 n-4)}},
\label{eqCorrEnBn}
\end{equation}
with limit
\begin{equation}
    \lim_{n \rightarrow \infty} \tcorr{E_n,B_n} = 0.
\label{eqCorrEnBnLim}
\end{equation}

\subsection{$I_n$ and $B_n$}
\label{calc:InBn}

We use eq.~\ref{eq:In_def} and results involving $L_n$ (eq.~\ref{eq:CovLnBn}) and $E_n$ (eq.~\ref{eqCovEnBn}) to obtain
\begin{align}
	\tcov{I_n, B_n} & = \Cov[L_n, B_n] - \tcov{E_n,B_n} \nonumber \\
	& = \frac{4(S_{3,n-1}n - S_{2,n-1} n + n - S_{3,n-1} - 1)}{n-1}, 
\label{eq:CovInBn2}
\end{align}
with limit
\begin{equation}\label{eq:CovInBnlim}
	\lim_{n\to\infty} \tcov{I_n,B_n} = 4 \zeta(3) + 4 - \frac{2 \pi^2}{3} \approx 2.22849.
\end{equation}

Dividing eq.~\ref{eq:CovInBn2} by the product of the square roots of eqs.~\ref{eq:var_In_def} and \ref{eq:var_Bn}, the approximate correlation is 
\begin{equation}\label{eq:CorrInBn}
	\tcorr{I_n,B_n} = \frac{\sqrt{2}(S_{3,n-1}n - S_{2,n-1}n + n -S_{3,n-1} - 1) n \sqrt{n-2}}{\sqrt{(n-1)[4 S_{1,n-1}+S_{2,n-1}(n-1) (n-2) -4(n-1)] (3S_{2,n-1}n^2 - 2S_{2,n-1}^2 n^2 + n^2 - 4S_{2,n-1}n + 3n-4)}},
\end{equation}
with limit
\begin{equation}\label{eq:CorrInBnlim}
	\lim_{n\to\infty} \tcorr{I_n,B_n} = \frac{\sqrt{6} [6 \zeta(3) + 6 - \pi^2]}{\pi \sqrt{18+9\pi^2-\pi^4}} \approx 0.84930.
\end{equation}


\section{Numerical and simulation-based analysis}
\label{secAnalysis}

\subsection{Analysis methods}

We examine the results of Section \ref{secTheoreticalResults} summarized in Tables \ref{tbl:covariances} and \ref{tbl:correlations} numerically and by coalescent simulation. For 13 of 15 covariances and correlations, the theoretical results are exact, and simulations merely verify that the mathematics has proceeded without error. For the covariances and correlations involving $(E_n,B_n)$ and $(I_n,B_n)$, the theoretical results are approximate, and the simulations assess the accuracy of the approximations.

We simulated the coalescent process for a series of values of $n$ beginning with $n=2$, at each value of $n$ performing 100,000 replicate simulations. To generate the simulated replicates, we employed \texttt{ms} \citep{mshudson}, using the command \texttt{ms n 100000 -T}, with \texttt{n} taken from $\{2, 3, \ldots, 50\}$. In the set of simulated replicates, we evaluated simulated covariances and correlation coeficients for pairs of quantities.

\begin{figure}[tbp]\centering
\includegraphics[width=0.6\textwidth, trim = 0 100 0 120, clip]{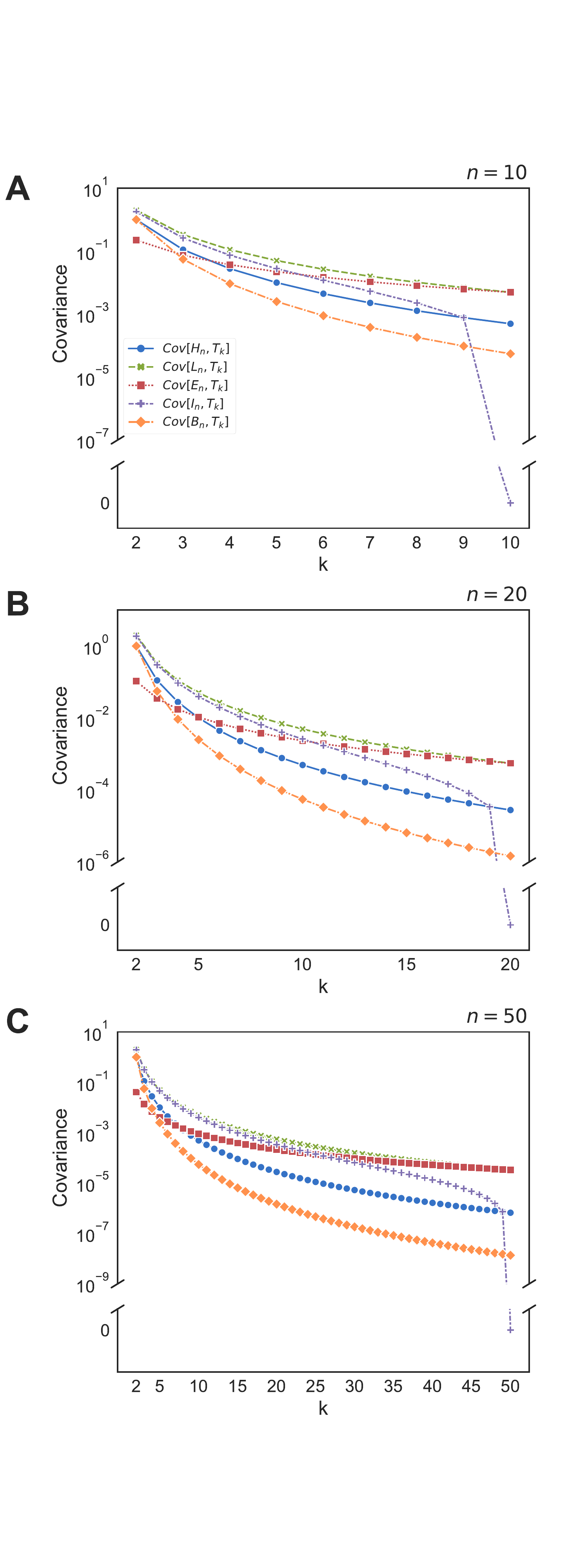}
\caption{Theoretical values of covariances $\Cov[X, T_k]$ for variables $X$ in $\{H_n,L_n,E_n,I_n,B_n\}$, plotted as functions of $k$ for $n=10$, $n=20$, and $n=50$. The plots appear on a logarithmic scale.}
\label{fig:Tk_cov_plots}
\end{figure}

We plot the mathematical results of Tables \ref{tbl:covariances} and \ref{tbl:correlations} together with simulation values in Figures \ref{fig:cov}-\ref{fig:Tk_corr_plots}. Figures \ref{fig:cov} and \ref{fig:Tk_cov_plots} show covariances of pairs of variables; Figures \ref{fig:corr} and \ref{fig:Tk_corr_plots}
show correlations. 

\subsection{Accuracy of approximations}

Figure \ref{fig:cov} shows the analytical and simulated covariances, and Figure \ref{fig:corr} shows the analytical and simulated correlations, for pairs of variables among $\{H_n,L_n,I_n,E_n,B_n\}$. For pairs of variables for which no approximations were needed in obtaining covariances---all except $(E_n,B_n)$ and $(I_n,B_n)$---the simulated and analytical values produce plots that are nearly indistinguishable.

For $(E_n,B_n)$ and $(I_n,B_n)$, the approximate and simulated correlations are close, but noticeably different (Figure \ref{fig:corr}); the mean absolute difference between the analytical and simulated values across choices of $n$ from 4 to 30 is $0.02458$ for $(E_n,B_n)$ and $0.01089$ for $(I_n,B_n)$. For covariance, which unlike the correlation coefficient is not standardized to lie in $[-1,1]$, the approximate and simulated values are quite close, with corresponding mean absolute deviations of $0.02372$ for $(E_n,B_n)$ and $0.03101$ for $(I_n,B_n)$.

\subsection{Properties of correlations}

\begin{figure}[tbp]\centering
\includegraphics[width=0.6\textwidth, trim = 0 100 0 120, clip]{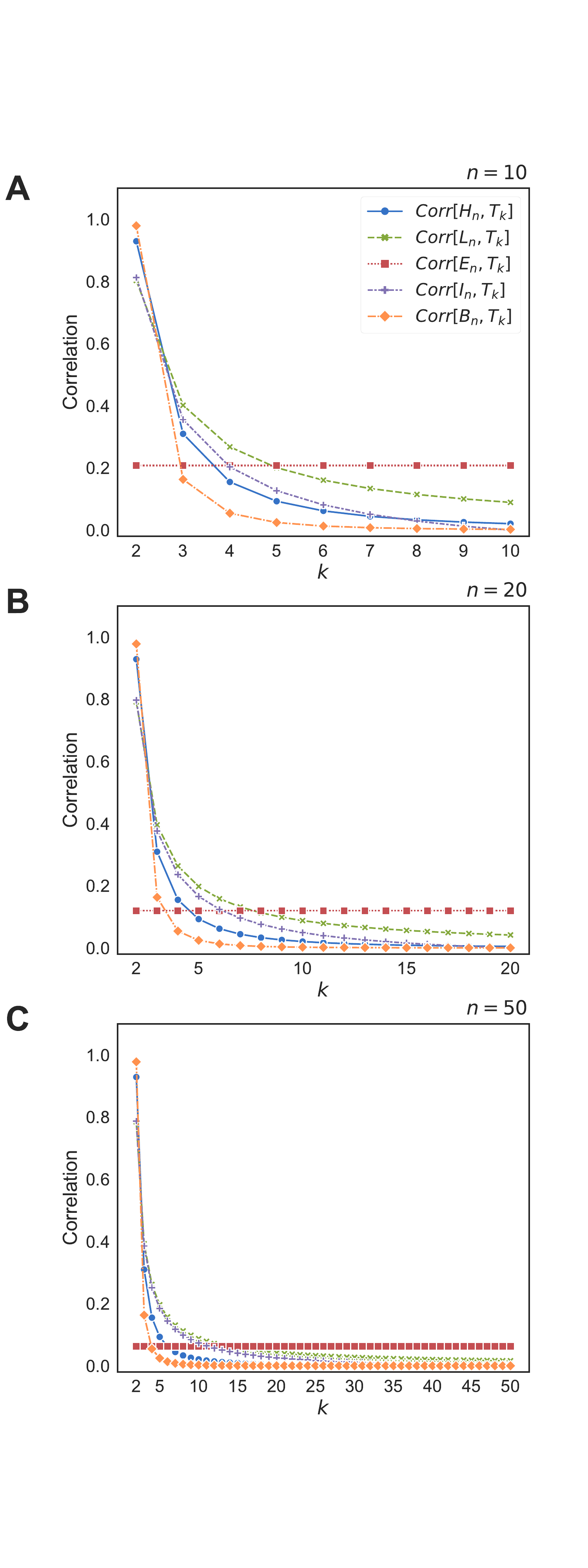}
\caption{Theoretical values of correlation coefficients $\Corr[X, T_k]$ for variables $X$ in $\{H_n,L_n,E_n,I_n,B_n\}$, plotted as functions of $k$ for $n=10$, $n=20$, and $n=50$.}
\label{fig:Tk_corr_plots}
\end{figure}

We observe that $H_n$, $L_n$, $I_n$, and $B_n$ all remain strongly correlated as $n$ increases, with the six limiting correlations among these four quantities lying between 0.84930 for $\Corr[L_n,B_n]$ and $\Corr[I_n,B_n]$ and 1 for $\Corr[L_n,I_n]$ (Table \ref{tbl:correlations}). The high limiting $\Corr[H_n,L_n]$ of approximately $0.93399$ reflects the strong influence of times $T_k$ with small $k$ on both $H_n$ and $L_n$ (Figures \ref{fig:Tk_cov_plots} and \ref{fig:Tk_corr_plots}). As $n$ increases, $\expect{I_n}$ increases without bound (eq.~\ref{eq:exp_In_def}), whereas $\expect{E_n}$ remains constant (eq.~\ref{eq:exp_En_def}); the contribution of $E_n$ to the total tree length $L_n$ becomes negligible, and $\Corr[L_n,I_n]$ approaches 1. $\Corr[H_n,I_n]$ has the same limiting value as $\Corr[H_n,L_n]$, and $H_n$, $L_n$, and $I_n$ all have limiting correlation 0 with $E_n$. Interestingly, although $H_n$ and $E_n$ have the same limiting expectation of 2, the limit of their correlation $\Corr[H_n,E_n]$ is 0.

The correlations of $H_n$, $L_n$, and $I_n$ with $B_n$, like their correlations with each other, are relatively high. $\Corr[H_n, B_n]$ is nearly constant in $n$, with limit approximately $0.97054$; both $H_n$ and $B_n$ are determined in large part by the $T_k$ with small $k$ (eqs.~\ref{eq:exp_Hn_def} and \ref{eq:exp_Bn_def}), so that little change occurs in the correlation as $n$ increases. Because $\Corr[H_n, B_n]$ is high and $\Corr[H_n,L_n]$ is also high, the constraint on a correlation $\Corr[Y,Z]$ given $\Corr[X,Y]$ and $\Corr[X,Z]$, or \citep[eq.~7.1]{wickens2014},
\begin{eqnarray}
\label{eqWickens}
\Corr[Y,Z] & \geq & \Corr[X,Y] \, \Corr[X,Z] - \sqrt{1-\Corr[X,Y]^2} \sqrt{1-\Corr[X,Z]^2} \\
\label{eqWickens2}
\Corr[Y,Z] & \leq & \Corr[X,Y] \, \Corr[X,Z] + \sqrt{1-\Corr[X,Y]^2} \sqrt{1-\Corr[X,Z]^2},
\end{eqnarray}
forces a high value for $\Corr[L_n,B_n]$ as well. In particular, placing $H_n,L_n,B_n$ in the roles of $X,Y,Z$, with $\lim_{n \rightarrow \infty} \Corr[H_n,L_n] \approx 0.93399$ and $\lim_{n \rightarrow \infty} \Corr[H_n,B_n] \approx 0.97054$, we obtain an interval $0.82037 \leq \lim_{n \rightarrow \infty} \Corr[L_n,B_n] \leq 0.99256$ from eqs.~\ref{eqWickens} and \ref{eqWickens2}; $\lim_{n\rightarrow \infty} \Corr[L_n,B_n] \approx 0.84930$ lies near its lower end.  Eqs.~\ref{eqWickens} and \ref{eqWickens2} similarly force a high value for $\lim_{n\rightarrow \infty} \Corr[I_n,B_n]$, using $H_n,I_n,B_n$ as $X,Y,Z$. 

Next, for correlations involving the $T_k$, we observe that for fixed $n$, as $k$ increases from 2 to $n$, $\Corr[H_n,T_k]$ decreases (Figure~\ref{fig:Tk_corr_plots}). At fixed $n$ and $k$, $\Corr[L_n,T_k]$ generally exceeds $\Corr[H_n,T_k]$; $k$ copies of the branch length $T_k$ contribute to tree length $L_n$ (eq.~\ref{eq:Ln_def}), whereas only one copy contributes to the tree height $H_n$ (eq.~\ref{eq:Hn_def}), giving rise to a greater value for the correlation of $T_k$ with $L_n$ than with $H_n$. For $k > 2$, $\Corr[B_n,T_k]$ is generally smaller than $\Corr[H_n,T_k]$; because $B_n$ is determined to a larger extent by $T_2$ than is $H_n$, the correlations of $B_n$ with $T_k$ for $k > 2$ are generally smaller. Finally, because tree length $L_n$ consists primarily of internal branches for large $n$, the correlation $\Corr[I_n,T_k]$ is similar to $\Corr[L_n,T_k]$ (Figure~\ref{fig:Tk_corr_plots}), approaching the same limit as $n \rightarrow \infty$ (Table~\ref{tbl:correlations}); the correlation of $E_n$ and $T_k$ is a constant that does not depend on $k$.

\section{Discussion}

\begin{figure}[!ht]
   \centering
   \includegraphics[width=\textwidth]{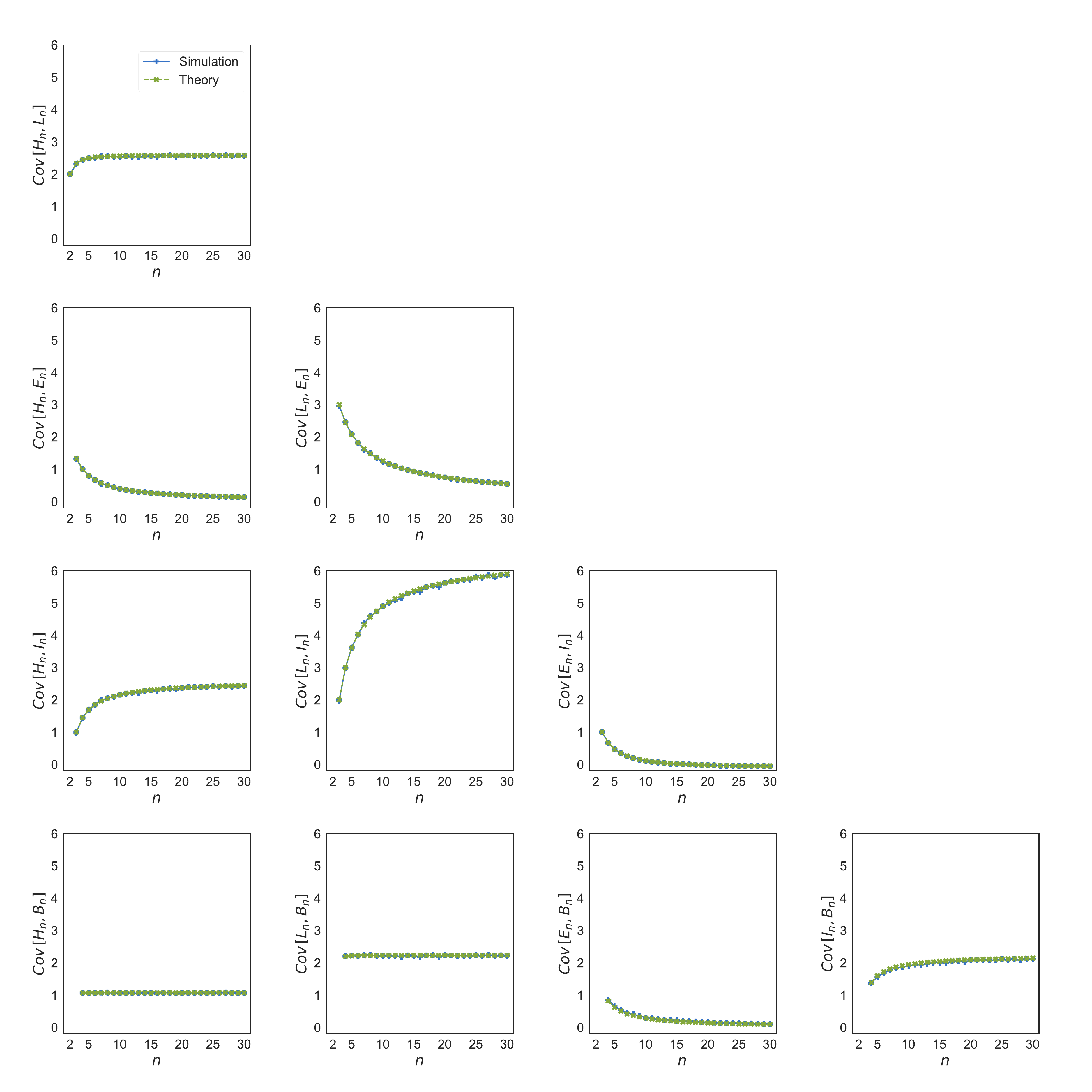}
   \caption{Simulated and theoretical values of covariances of pairs of variables, plotted as functions of sample size $n$. Expressions for theoretical values are taken from Table~\ref{tbl:covariances}.}
   \label{fig:cov}
\end{figure}

\begin{figure}[!ht]
   \centering
   \includegraphics[width=\textwidth]{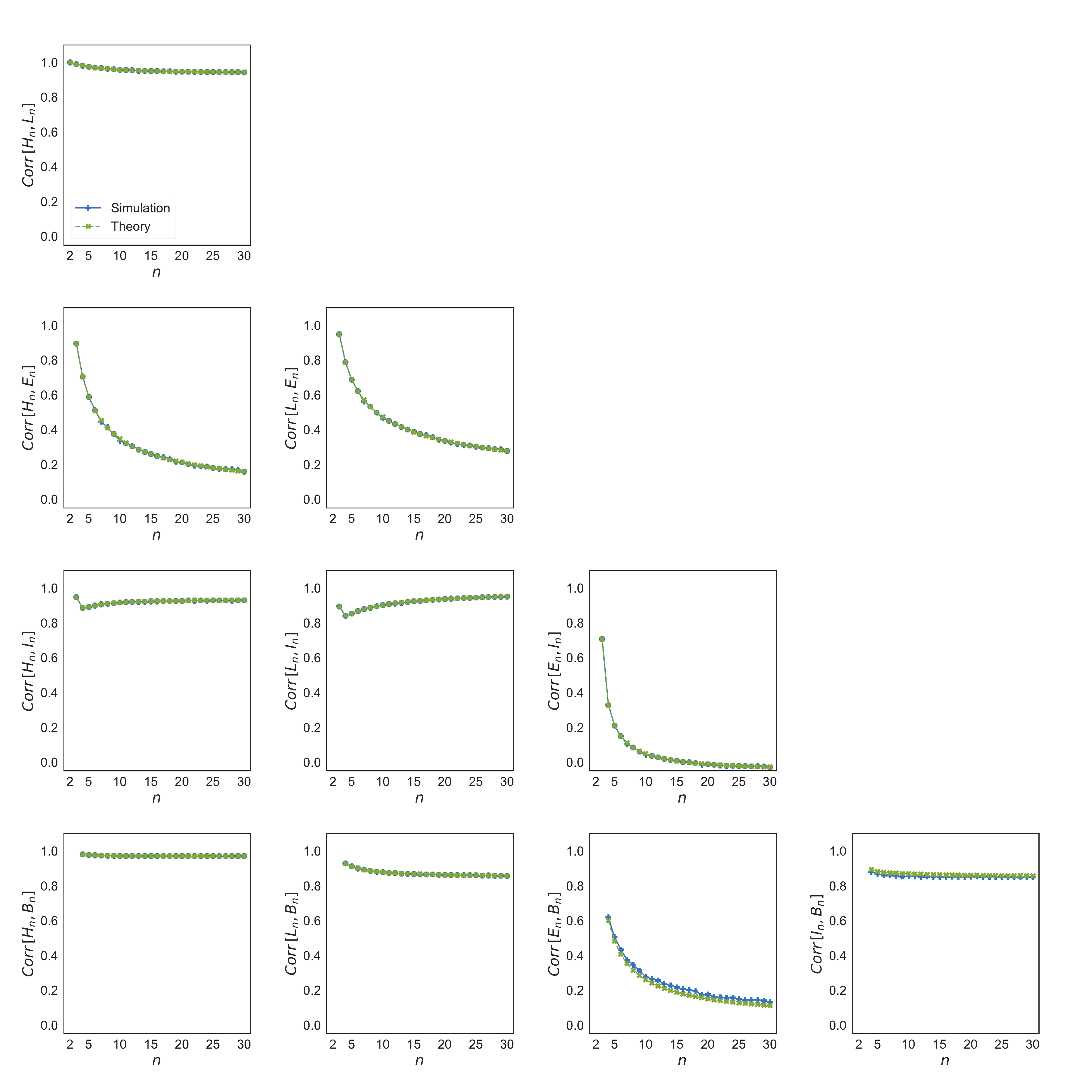}
   \caption{Simulated and theoretical values of correlation coefficients of pairs of variables, plotted as functions of sample size $n$. Expressions for theoretical values are taken from Table~\ref{tbl:correlations}.}
   \label{fig:corr}
\end{figure}

We have examined relationships between pairs of tree features under the coalescent model by deriving expressions for their covariances and correlation coefficients (Tables \ref{tbl:covariances} and \ref{tbl:correlations}). For 13 of 15 pairs examined, we obtained exact expressions for the covariances and correlation coefficients, and for the remaining two pairs, we obtained quantities observed in simulations to closely approximate the desired quantities (Figures \ref{fig:cov} and \ref{fig:corr}). The results provide a compendium of basic relationships among coalescent tree features, contributing to a more precise understanding of the way in which the properties of coalescent trees relate to each other.

In most cases, the covariances have relatively simple expressions, comparable to the simplicity of most expressions for expectations and variances (Table \ref{tbl:covariances}). Expressions for the correlation coefficients are somewhat more complex, in many cases with $n \rightarrow \infty$ limits that contain terms resulting from the limit $\sum_{k=1}^\infty 1/k^2 = \pi^2/6$.

Numerically, we obtain tight correlations between $H_n$, $L_n$, $I_n$, and $B_n$ as $n$ grows large, with all of these quantities possessing limiting correlations of $0.84930$ or greater (Table \ref{tbl:correlations}). In the limit, $L_n$ and $I_n$ are perfectly correlated, and all limiting correlations of other quantities with $I_n$ are equal to their corresponding correlations with $L_n$. Decreasing correlations are observed for $H_n$, $L_n$, $I_n$, and $B_n$ with $E_n$, with limits of 0 observed in all cases (Table 2). Although $H_n$ and $E_n$ both have limiting expectation 2 (eqs.~\ref{eq:exp_Hn_def} and \ref{eq:exp_En_def}), their limiting correlation coefficient is 0. The correlations among $H_n$, $L_n$, and $B_n$ are all large; however, the limiting correlation for $(L_n,B_n)$ is near the lower end of the interval suggested by the larger limiting correlations for $(H_n,L_n)$ and $(H_n,B_n)$ (eqs.~\ref{eqWickens} and \ref{eqWickens2}). This result suggests that $L_n$ and $B_n$ capture relatively distinct features of coalescent trees in relation to the constraints placed on a pair of correlated variables that are each highly correlated with a third variable ($H_n$). A similar observation can be made concerning $I_n$ and $B_n$, as $L_n$ and $I_n$ are asymptotically fully correlated.

\tcr{Although tree properties such as $H_n$, $L_n$, $E_n$, $I_n$, and $B_n$ are not themselves observable in genetic sequences, interest in these quantities} arises in part from their relationship to statistical tests that assess the fit of the coalescent model to data on genetic variation. Features of tree shape underlie predictions of the coalescent regarding allele frequencies; in particular, tree properties contribute to predictions for the unfolded site-frequency spectrum (SFS) of a genomic region, the vector that for a sample of size $n$ tabulates how many variable (biallelic) sites in the region possess allele frequencies $1/n, 2/n, \ldots, (n-1)/n$ for the derived allele \citep[e.g.][]{fu1995,ferretti2017}. Test statistics then assess agreement of site-frequency spectra with the predictions \tcr{\citep[e.g.][]{zeng_fu2006, achaz2009,ferretti2010,ronen2013,fu2021}}, so that correlations among statistics emphasizing different aspects of site-frequency spectra emerge from dependence on correlated tree features. In this context, further understanding of correlations among tree properties can assist in understanding the \tcr{joint} behavior of SFS-based tests of the coalescent model.

Our computations augment earlier calculations concerning quantities associated with coalescent trees. The pairs $(H_n,L_n)$ \citep{arbisser2018joint} and $(L_n,E_n)$ and $(E_n,I_n)$ \citep{fu1993statistical} have been studied in detail. \tcr{Results for pairs $(H_n,T_k)$, $(L_n,T_k)$, and $(L_n,I_n)$ follow trivially from the derivations and results of \cite{arbisser2018joint} and \cite{fu1993statistical}, but were not highlighted in those studies. Results for pairs $(H_n,E_n)$, $(H_n,I_n)$, $(E_n,T_k)$, and $(I_n,T_k)$ follow from derivations similar to those of \cite{fu1993statistical}, but to our knowledge, they have not been previously reported.}

\tcr{The least-studied of the variables we consider, $B_n$, was introduced by \cite{uyenoyama1997} in the context of balancing selection and self-incompatibility alleles in plants. Under balancing selection, the mean $B_n$ of the two basal branches is expected to be long in relation to the tree length $L_n$, so that $2B_n/L_n$ predicts the fraction of segregating sites that distinguish two long-separated sets of lineages. For $B_n$, which gives a portion of the height $H_n$---but which, unlike $H_n$, is obtained from a sum with a random length---we derived the variance} (eq.~\ref{eq:var_Bn}), as well as exact covariances and correlations with $H_n$, $L_n$, and $T_k$ and approximate covariances and correlations with $E_n$ and $I_n$. Several studies have extended the work of \cite{fu1993statistical} on features of the external and internal branch lengths \citep{blum2005, caliebe2007, janson2011, dahmer2015internal, dahmer2017total, disanto2020}; it may be possible to seek exact rather than approximate covariances and correlations for $(E_n,B_n)$ and $(I_n,B_n)$ by building on these studies.

\tcr{When examining joint distributions of $H_n$ and $L_n$, \cite{arbisser2018joint} used computations of the expectations and variances of $H_n$ and $L_n$ and the covariance of $H_n$ and $L_n$ to obtain approximations for the expectation and variance of $H_n/L_n$. Following the approach of \cite{arbisser2018joint}, our results could be used to obtain similar approximate expressions for expectations and variances of ratios of additional pairs.}

\vskip .3cm
\noindent {\bf Acknowledgments.} We thank Yun-Xin Fu and John Wakeley for comments on the manuscript. We acknowledge support from NIH grants R01 GM131404 and R01 HG005855 and NSF grant BCS-2116322.

{\footnotesize
\bibliographystyle{tpb}
\bibliography{sources.bib}
}

\newpage
\linespread{1}

\newgeometry{left=0.5in,right=0.5in,top=0.5in,bottom=0.5in}
\begin{landscape}
{\renewcommand{\arraystretch}{1.7}
\begin{table}[ht]
	\centering
	\begin{tabular}{lllp{0.3\linewidth}}
	\toprule
 $(X_n,Y_n)$ & $\Cov[X_n,Y_n]$ & $\lim_{n \rightarrow \infty} \Cov[X_n,Y_n]$ & Reference \\
	\midrule
	$H_n$, $T_k$ & $\frac{4}{k^2(k-1)^2}$ & $\frac{4}{k^2(k-1)^2}$ & \ref{calc:HnTk}, eq.~\ref{eq:CovTkHn} \\
	$H_n$, $L_n$ & $4S_{2,n-1} - 4 + \frac4n$ & $\frac{2\pi^2}{3} - 4 \approx 2.57974$ & \ref{calc:HnLn}, eqs.~\ref{eq:CovHnLn}, \ref{eq:CovHnLnlim} \\
	$H_n$, $E_n$ & $\frac{4}{n}$ & 0 & \ref{calc:HnEn}, eqs.~\ref{eq:CovHnEn}, \ref{eq:CovHnEnlim}  \\
	$H_n$, $I_n$ & $4 S_{2,n-1}-4$ & $\frac{2\pi^2}{3} -4 \approx 2.57974$ & \ref{calc:HnIn}, eqs.~\ref{eq:CovHnIn}, \ref{eq:CovHnInlim} \\
	$H_n$, $B_n$ & $\frac{4[S_{3,n-1}n^2-3S_{2,n-1}n^2+(n-1)(4n+1)]}{n^2}$ & $4\zeta(3) + 16 - 2\pi^2 \approx 1.06902$ & \ref{calc:HnBn}, eqs.~\ref{eq:CovHnBn}, \ref{eq:CovHnBnlim} \\
	$L_n$, $T_k$ & $\frac{4}{k(k-1)^2}$ & $\frac{4}{k(k-1)^2 }$ & \ref{calc:LnTk}, eq.~\ref{eq:CovTkLn} \\
	$L_n$, $E_n$ & $\frac{4 S_{1,n-1}}{n-1}$ & 0 & \ref{calc:LnEn}, eqs.~\ref{eq:CovLnEn}, \ref{eq:CovLnEnlim} \\
	$L_n$, $I_n$ & $4 S_{2,n-1}-\frac{4 S_{1,n-1}}{n-1}$ & $\frac{2\pi^2}{3} \approx 6.57974$ & \ref{calc:LnIn}, eqs.~\ref{eq:CovLnIn}, \ref{eq:CovLnInlim} \\
	$L_n$, $B_n$ & $\frac{4[S_{3,n-1}n-S_{2,n-1}n+n-1]}{n}$ & $4\zeta(3) + 4 - \frac{2\pi^2}{3} \approx 2.22849$ & \ref{calc:LnBn}, eqs.~\ref{eq:CovLnBn}, \ref{eq:CovLnBnlim} \\
	$E_n$, $T_k$ & $\frac{4}{k(k-1)(n-1)}$ & 0 & \ref{calc:EnTk}, eqs.~\ref{eq:CovEnTk}, \ref{eq:CovEnTklim} \\
	$E_n$, $I_n$ & $\frac{4S_{1,n-1}}{n-1} - \frac{8S_{1,n-1}n}{(n-1)(n-2)} + \frac{16}{n-2}$  & 0 & \ref{calc:EnIn}, eqs.~\ref{eq:CovEnIn}, \ref{eq:CovEnInlim} \\
	$E_n$, $B_n$ & $\frac{4 \left(S_{2,n-1}n-n+1\right)}{n(n-1)}$ & 0 & \ref{calc:EnBn}, eqs.~\ref{eqCovEnBn}, \ref{eqCovEnBnLim} \\
	$I_n$, $T_k$ & $\frac{4 (n-k)}{k(k-1)^2 (n-1)}$ & $\frac{4}{k(k-1)^2}$ & \ref{calc:InTk}, eqs.~\ref{eq:CovInTk}, \ref{eq:CovInTklim} \\
	$I_n$, $B_n$ & $\frac{4(S_{3,n-1}n - S_{2,n-1} n + n - S_{3,n-1} - 1)}{n-1}$
    & $4\zeta(3) + 4 - \frac{2\pi^2}{3} \approx 2.22849$ & \ref{calc:InBn}, eqs.~\ref{eq:CovInBn2}, \ref{eq:CovInBnlim} \\
	$B_n$, $T_k$ & $ \frac{4}{k^2(k-1)^3}$ & $ \frac{4}{k^2(k-1)^3}$ & \ref{calc:BnTk}, eq.~\ref{eq:CovBnTk} \\
	\bottomrule
	\end{tabular}
	\caption{Covariances of pairs of variables. Expressions involving $E_n$ or $I_n$ apply for $n \geq 3$ and expressions involving $B_n$ apply for $n \geq 4$.}\label{tbl:covariances}
\end{table}
}
\clearpage
{\renewcommand{\arraystretch}{1.7}
\begin{table}[ht]
	\centering
	\begin{tabular}{lllp{0.2\linewidth}}
	\toprule
$(X_n,Y_n)$ & $\Corr[X_n,Y_n]$ & $\lim_{n \rightarrow \infty} \Corr[X_n,Y_n]$ & Reference \\
	\midrule
	$H_n$, $T_k$ & $\frac{n}{\sqrt{2 n^2 S_{2,n} - 3n^2 + 2n -1}} \frac{1}{k(k-1)} $ & $\frac{\sqrt{3}}{\sqrt{\pi^2-9}} \frac{1}{k(k-1)} \approx \frac{1.85738}{k(k-1)} $ & \ref{calc:HnTk}, eqs.~\ref{eq:CorrTkHn}, \ref{eq:CorrTkHnlim} \\
	$H_n$, $L_n$ & $\frac{S_{2,n-1}n-n+1}{\sqrt{S_{2,n-1} \left(2S_{2,n}n^2-3n^2+2n-1 \right)}}$ & $\frac{\pi^2-6}{\pi \sqrt{2(\pi^2-9)}} \approx 0.93399$ & \ref{calc:HnLn}, eqs.~\ref{eq:corr_Hn_Ln}, \ref{eq:corr_Hn_Ln_limit} \\
	$H_n$, $E_n$ & $\frac{\sqrt{(n-1)(n-2)}}{\sqrt{2( 2 S_{2,n}n^2 - 3n^2 + 2n - 1) \left(S_{1,n-1}n-2 n+2\right)}}$ & 0 & \ref{calc:HnEn}, eqs.~\ref{eq:CorrHnEn}, \ref{eq:CorrHnEnlim} \\
	$H_n$, $I_n$ & $\frac{ \left(S_{2,n-1}-1\right)n\sqrt{(n-1)(n-2)}} {\left(2S_{2,n} n^2 -3 n^2+2 n-1\right) \left[4 S_{1,n-1}+S_{2,n-1}(n-1)(n-2) -4(n-1)\right]}$ &  $\frac{\pi^2-6}{\pi \sqrt{2(\pi^2-9)}} \approx 0.93399$ & \ref{calc:HnIn}, eqs.~\ref{eq:CorrHnIn}, \ref{eq:CorrHnInlim} \\
	$H_n$, $B_n$ & $\frac{\sqrt{2}[S_{3,n-1}n^2 - 3S_{2,n-1}n^2 + (n-1)(4n+1)]}{ \sqrt{(2S_{2,n}n^2-3n^2+2n-1) (3S_{2,n-1}n^2-2S_{2,n-1}^2n^2+ n^2-4S_{2,n-1}n+3n-4})}$ & $\frac{3\sqrt{3} [2\zeta(3) + 8 - \pi^2]}{\sqrt{(\pi^2-9) 
	(18+9\pi^2-\pi^4)}} \approx 0.97054$ & \ref{calc:HnBn}, eqs.~\ref{eq:CorrHnBn}, \ref{eq:CorrHnBnlim} \\
	$L_n$, $T_k$ & $\frac{1}{\sqrt{S_{2,n-1}}} \frac{1}{k-1}$ & $\frac{\sqrt{6}}{\pi} \frac{1}{k-1} \approx \frac{0.77970}{k-1}$ & \ref{calc:LnTk}, eqs.~\ref{eq:CorrTkLn}, \ref{eq:CorrTkLnlim} \\
	$L_n$, $E_n$ & $\frac{S_{1,n-1}\sqrt{n-2}}{\sqrt{2S_{2,n-1} (n-1) \left(S_{1,n-1}n-2 n+2\right)}}$ & 0 & \ref{calc:LnEn}, eqs.~\ref{eq:CorrLnEn}, \ref{eq:CorrLnEnlim} \\
	$L_n$, $I_n$ & $\frac{[S_{2,n-1}(n-1)-S_{1,n-1}]\sqrt{n-2}}{(n-1) \sqrt{S_{2,n-1} \left[4 S_{1,n-1}+S_{2,n-1}(n-1)(n-2) -4(n-1)\right]}}$ & 1 & \ref{calc:LnIn}, eqs.~\ref{eq:CorrLnIn}, \ref{eq:CorrLnInlim} \\
	$L_n$, $B_n$ & $\frac{\sqrt{2}(S_{3,n-1}n - S_{2,n-1}n + n-1)} { \sqrt{S_{2,n-1} (3S_{2,n-1}n^2-2S_{2,n-1}^2n^2+n^2-4S_{2,n-1}n+3n-4})}$ & $\frac{\sqrt{6} [6\zeta(3) + 6 - \pi^2]}{\pi \sqrt{18+9\pi^2-\pi^4}}  \approx 0.84930$ & \ref{calc:LnBn}, eqs.~\ref{eq:CorrLnBn}, \ref{eq:CorrLnBnlim} \\
	$E_n$, $T_k$ & $\frac{\sqrt{n-2}}{\sqrt{2(n-1)(S_{1,n-1}n-2n+2)}}$ & 0 &  \ref{calc:EnTk}, eqs.~\ref{eq:CorrEnTk}, \ref{eq:CorrEnTklim}\\
	$E_n$, $I_n$ & $\frac{4(n-1) -  S_{1,n-1}(n+2)}{\sqrt{2 \left( S_{1,n-1}n-2 n+2\right) \left[4 S_{1,n-1}+ S_{2,n-1}(n-1)(n-2) - 4(n-1)\right]}}$ & 0 & \ref{calc:EnIn}, eqs.~\ref{eq:CorrEnIn}, \ref{eq:CorrEnInlim} \\
	$E_n$, $B_n$ & $\frac{(S_{2, n-1}n-n+1)\sqrt{n-2}}
    {\sqrt{ (n-1) (S_{1, n-1}n -2 n+2) (3S_{2, n-1}n^2 -2S_{2,n-1}^2 n^2 +n^2 - 4S_{2, n-1} n +3 n-4)}}$
	& 0 & \ref{calc:EnBn}, eqs.~\ref{eqCorrEnBn}, \ref{eqCorrEnBnLim} \\
	$I_n$, $T_k$ & $\frac{(n-k)\sqrt{n-2}} {(k-1)\sqrt{(n-1)\left[4 S_{1,n-1}+S_{2,n-1}(n-1)(n-2) -4(n-1) \right]}}$ & $\frac{\sqrt{6}}{\pi} \frac{1}{k-1} \approx \frac{0.77970}{k-1}$ & \ref{calc:InTk}, eqs.~\ref{eq:CorrInTk}, \ref{eq:CorrInTklim} \\
	$I_n$, $B_n$ &  $\frac{\sqrt{2}(S_{3,n-1}n - S_{2,n-1}n + n -S_{3,n-1} - 1) n \sqrt{n-2}}{\sqrt{(n-1)[4 S_{1,n-1}+S_{2,n-1}(n-1) (n-2) -4(n-1)] (3S_{2,n-1}n^2 - 2S_{2,n-1}^2 n^2 + n^2 - 4S_{2,n-1}n + 3n-4)}}$ & $\frac{\sqrt{6} [6\zeta(3) + 6 - \pi^2]}{\pi \sqrt{18+9\pi^2-\pi^4}} \approx 0.84930$ & \ref{calc:InBn}, eqs.~\ref{eq:CorrInBn}, \ref{eq:CorrInBnlim} \\
	$B_n$, $T_k$ & $\frac{\sqrt{2} n}{ \sqrt{3S_{2,n-1} n^2 - 2 S_{2,n-1}^2 n^2 + n^2 -4S_{2,n-1}n +3n-4}} \frac{1}{k(k-1)^2}$  & $\frac{6}{\sqrt{18+9 \pi ^2-\pi ^4}} \frac{1}{k(k-1)^2} \approx \frac{1.95518}{k(k-1)^2}$ & \ref{calc:BnTk}, eqs.~\ref{eq:CorrBnTk}, \ref{eq:CorrBnTklim} \\
	\bottomrule
	\end{tabular}
	\caption{Correlation coefficients of pairs of variables. Expressions involving $E_n$ or $I_n$ apply for $n \geq 3$ and expressions involving $B_n$ apply for $n \geq 4$. }\label{tbl:correlations}
\end{table}
}
\end{landscape}
\restoregeometry

\clearpage
\clearpage
\clearpage
\clearpage

\end{document}